\documentclass[%
 reprint,
 amsmath,amssymb,
 aps,
]{revtex4-2}

\usepackage{graphicx}
\usepackage{subcaption}  
\usepackage{caption}
\usepackage{ragged2e} 

\usepackage{subfig}
\usepackage{afterpage}
\usepackage{dcolumn}
\usepackage{bm}

\begin{document}

\preprint{APS/123-QED}

\title{Pseudogap in \(Sr_{2-x}La_{x}IrO_{4}\): Gor'kov-Teitel'baum thermal activation model}

 \author{
  \hspace{5cm}
  Jalaja Pandya\textsuperscript{1},
  Devarshi Dave\textsuperscript{2},
  Navinder Singh\textsuperscript{1}
  \vspace{0.3cm} 
  \newline 
  \textsuperscript{1}Physical Research Laboratory, Ahmedabad, India, PIN: 380009.\\
  \textsuperscript{2}Chaitanya School, Gandhinagar
}  


\date{\today}

\begin{abstract}
Recently, Hall effect measurements are done on Lanthanum doped Strontium Iridate \(Sr_{2-x}La_{x}IrO_{4}\) which is 5d analogue of cuprates \cite{hsu2023carrier}. Hall effect measurements show that the effective carrier density \(n_{H}\) exhibits a crossover from \(n_{H} \sim x\) to \(n_{H} \sim 1+x\) near \(x\simeq 0.16\) . This is very similar to that found in cuprates around $p\simeq0.19$. It is proposed that a pseudogap (PG) state in \(Sr_{2-x}La_{x}IrO_{4}\) exists and is ending at \(x\simeq 0.16\) \cite{hsu2023carrier}. However, PG boundary (in doping-temperature phase diagram) remains unknown. In this work, we apply a very successfull the Gor'kov-Teitel'baum Thermal Activation (GTTA) model to \(Sr_{2-x}La_{x}IrO_{4}\) and obtain its PG phase boundary and draw an updated phase diagram. Our results agree with previously known signatures of PG phase in this system \cite{seo2017infrared,de2015collapse}. Using results from GTTA model we also obtain the evolution of "Fermi arcs" in this system as a function 
temperature for doping concentration $x\simeq0.08$ which are in qualitative agreement with the reported results.
\end{abstract}

\maketitle


\section{\label{sec:level1}Introduction}

The transition-metal oxides have gained considerable attention in the last few decades because of their diverse physical properties including high-\(T_{c}\) superconductivity \cite{imada1998metal,maekawa2004physics,cox2010transition,anderson2011personal,anderson2004physics,lee2006doping,barivsic2013universal,proust2019remarkable,singh2021leading}. The 3d transition metal copper oxides are the most extensively studied materials in which high-\(T_{c}\) superconductivity is realized by hole/electron doping of the Mott insulator  Another striking characteristic of these hole-doped cuprates is the presence of a "Pseudogap (PG)" phase in the underdoped regime at normal temperature (much higher than \(T_{c}\)) \cite{timusk1999pseudogap,ayres2022superfluid,norman2005pseudogap,tallon2001doping,tallon2023thermodynamics}. The PG phase boundary (more precisely a crossover) starts at a very high temperature ($\backsim$ 1000 K) near the Mott insulating phase and decreases linearly with increasing doping. It closes at a critical doping of \(p*\simeq0.19\), just above optimal doping regime \cite{tallon2020locating,taillefer2016change,gor2015two,taillefer2010scattering,michon2019thermodynamic,collignon2017fermi}. 

Strontium Iridate \(Sr_{2}IrO_{4}\) is a spin-orbit coupled 5d Mott insulator which shows similarities with cuprates. From simple valence counting arguments, strontium has valency: [Kr]\(5s^{2}\), iridium has valency: [Xe]\(4f^{14}5d^{7}6s^{2}\) and oxygen has valency: [He]\(2s^{2}2p^{4}\). In the ionic case scenario, the nominal valency state of \(Ir^{4+}\) can be estimated from \(+2\times2+x-2\times4=0\) i.e. \(x=+4\), as nominal valence of Sr and O will be \(Sr^{2+}\) and \(O^{2-}\) respectively. \(I^{4+}\) has the configuration \([Xe]5d^{5}\). Thus, out of 5 electrons one electron remains unpaired in the 5d shell, which exhibits magnetism and Mott insulation \cite{bertinshaw2019square}. A Canted AntiFerroMagnetic phase (C-AFM) is observed in the material below 230 K \cite{chen2015influence}. Under electron doping emergence of pseudogap and "Fermi arcs" are observed in the compound, resembling the hole-doped cuprates \cite{bertinshaw2019square}.

Several experimental studies have been reported in literature for electron doped \(Sr_{2}IrO_{4}\) to understand the character of the electronic phases of this material. In 2014, Kim et al used a surface doping technique for deposition of potassium atoms on \(Sr_{2}IrO_{4}\) and studied the evolution of the phenomenon of "Fermi arcs" with varying surface coverage and temperature \cite{kim2014fermi}. A decrease in the anti-nodal gap and an increase in "Fermi arc" lengths was seen with the increase in the surface coverage. A similar Angle-resolved photo-emission spectroscopy (ARPES) analysis for 7\% electron doped \(Sr_{2}IrO_{4}\) has been reported, wherein the variation of PG magnitude with temperature is studied \cite{kim2016observation}. Electron doping in \(Sr_{2}IrO_{4}\) can also be done by substituting \(Sr^{2+}\) by \(La^{3+}\) in the parent material. An evidence of a pseudogap was observed for \(Sr_{2-x}La_{x}IrO_{4}\) with \(x=0.10\) from the ARPES measurements \cite{de2015collapse}. Li et al performed ARPES and in-plane resistivity analysis for \(Sr_{2-0.30}La_{0.30}IrO_{4}\) thin films where they observed that although the system was insulating, the resistivity was suppressed for the doped material \cite{li2015tuning}. The electronic response of \(IrO_{2}\) plane for \(Sr_{2-0.134}La_{0.134}IrO_{4}\) using infrared spectroscopy revealed a PG of magnitude $\sim$17meV \cite{seo2017infrared}.\

Recently, Hsu et al carried out Hall-effect and specific heat analysis for \(Sr_{2-x}La_{x}IrO_{4}\) at several dopings \cite{hsu2023carrier}. They report variation in the Hall-coefficients of \(Sr_{2-x}La_{x}IrO_{4}\) for \(0<x<0.20\) as a function of temperature (T). The variation of \(C/T\) (C=specific heat) as a function of \(T^{2}\) is also reported. A dramatic change in the carrier density at around \(x\simeq0.16\) and a divergence in the extrapolated (T=0K) heat capacity indicates the presence of critical doping for PG opening below x $\leq$ 0.16.  Measured Hall number density indicates a "sudden" increase in carrier density from \(n_{H} \sim x\) to \(n_{H} \sim 1+x\) around that doping. Nevertheless, PG boundary is still unknown, which is addressed in this work.\   

The Gorkov-Teitelbaum Thermal Activation (GTTA) model was introduced by Lev P. Gor’kov and Gregory B. Teitel’baum in 2006 \cite{PhysRevLett.97.247003}. It is a phenomenological model for number of charge carriers which was derived from the Hall effect data of \(La_{2-x}Sr_{x}CuO_{4}\) \cite{PhysRevLett.97.247003}. The experimental Hall effect data of \(La_{2-x}Sr_{x}CuO_{4}\) can be rationalized by the GTTA model as follows:
\begin{equation}
    R_{H}=\frac{1}{n_{Hall}e}
    \label{GTTA}
\end{equation}
\begin{equation}
    n_{Hall}(x,T)=n_{0}(x)+n_{1}(x)e^{[-\Delta(x)/T]}
    \label{GTTA}
\end{equation}
The first term \(n_{0}(x)\), is the temperature independent term which incorporates the carrier density due to external hole doping. The second term is temperature dependent term which can be interpreted as thermal activation contribution of carriers above the gap. \(n_{0}(x)\) varies linearly with $x$ for low doping (up to \(x=0.07\)). For \(x>0.07\), \(n_{0}(x)\)  evolves non-linearly with $x$. \(n_{1}(x)\simeq2.8\) (roughly constant) for doping concentrations (\(x<0.19\)) and drops abruptly for  doping concentrations (\(x>0.19\)). $\Delta(x)$ is referred as the activation energy and corresponds to the energy difference between the "Fermi arc" and the band bottom for \(La_{2-x}Sr_{x}CuO_{4}\) \cite{PhysRevLett.97.247003}. The extracted $\Delta(x)$ decreases linearly with increase in \(x\) till \(x=0.20\). The \(\Delta(x)\) values are in agreement with ARPES results. 

We applied this very successful model (GTTA) to \(Sr_{2-x}La_{x}IrO_{4}\) and we have drawn an updated phase diagram (refer to Section 4) for the system. Reasons behind applying GTTA to \(Sr_{2-x}La_{x}IrO_{4}\) are given in section 2. We observe that at x=0.036, $\Delta$$\simeq$1200 K, then it linearly decreases with increasing doping and ends at \(x\simeq0.16\). This is quite similar to that found in cuprates. Using the carrier density obtained from GTTA model, the evolution of the "Fermi arcs" for doping concentration \(x\simeq0.08\) at temperature 70K and 100K is also obtained (refer to section 3).     

This paper is organized as follows: In section 2 we present our analysis of the Hall effect data using the GTTA model. In section 3 we present the development "Fermi arcs" with increasing temperature and for a given doping concentration ($x\simeq0.08$) for which the data is available. We conclude all our findings in the section 4. 

\section{Analyzing the Hall effect data using the GTTA model}
As discussed in the previous section, GTTA model is proven to be very successful for $La_{2-x}Sr_{x}CuO_{4}$ system \cite{PhysRevLett.97.247003}. Gor’kov and Teitel’baum calculated the doping dependent parameters viz. \(n_{0}(x), n_{1}\) and $\Delta(x)$ in Eq.2 using the experimental Hall effect data available in the literature \cite{ono2007strong,ando2004evolution}. The calculated $\Delta(x)$ agreed well with the ARPES experiments \cite{yoshida2003metallic,yoshida2006systematic}. \ 
\    

\begin{figure}[h]
  \centering
  \includegraphics[width=1.0\linewidth]{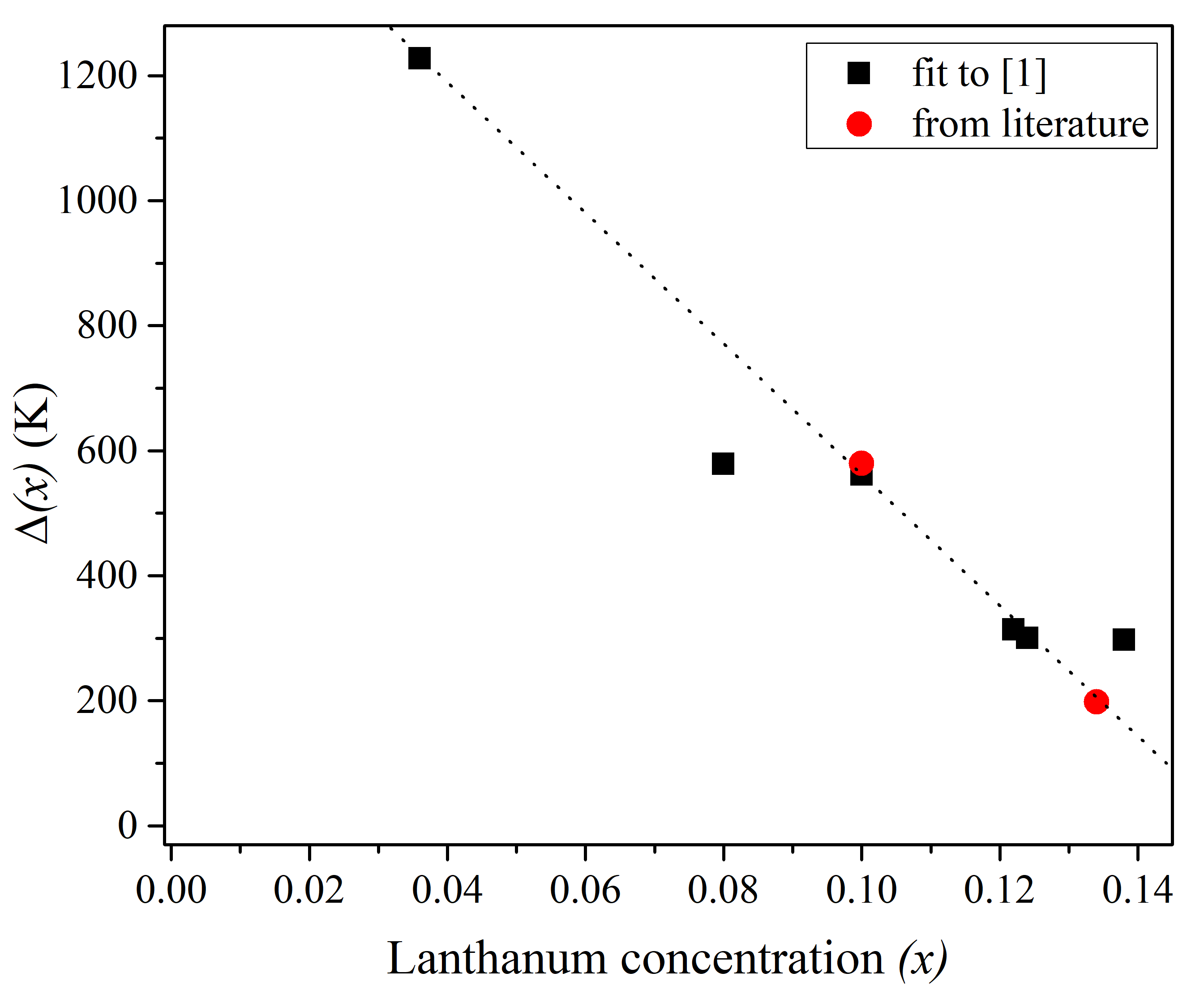}
  \caption{\justifying The activation energy $\Delta(x)$ obtained by fitting Eq. 2 \cite{PhysRevLett.97.247003} to the experimental Hall coefficient data for \(Sr_{2-x}La_{x}IrO_{4}\) for different doping concentrations extracted from Fig.1 of \cite{hsu2023carrier} and by private communications with the authors of \cite{hsu2023carrier}. Red solid circles represents experimental PG values obtained in \cite{seo2017infrared,de2015collapse}.}
  \label{fig:1}
\end{figure}

The crystal structure as well as the electronic structure (one unpaired electron in d orbitals) of $Sr_{2}IrO_{4}$ compound is similar to that of $La_{2}CuO_{4}$. Signatures of PG in \(Sr_{2-x}La_{x}IrO_{4}\) for some doping concentrations have been observed by ARPES measurements \cite{de2015collapse,kim2016observation} similar to $La_{2-x}Sr_{x}CuO_{4}$. The temperature dependence of the Hall coefficient for \(Sr_{2-x}La_{x}IrO_{4}\) is also similar to that of $La_{2-x}Sr_{x}CuO_{4}$.  Keeping in mind these strong similarities it is reasonable to apply the GTTA model to \(Sr_{2-x}La_{x}IrO_{4}\) system. We also obtain a-posteriori justification of the application of GTTA model as extracted PG agrees reasonably well with some pre-existing experimental data \cite{de2015collapse,seo2017infrared}.
\

\begin{figure}[h]
  \begin{subfigure}[b]{0.5\columnwidth}
        \includegraphics[width=\textwidth]{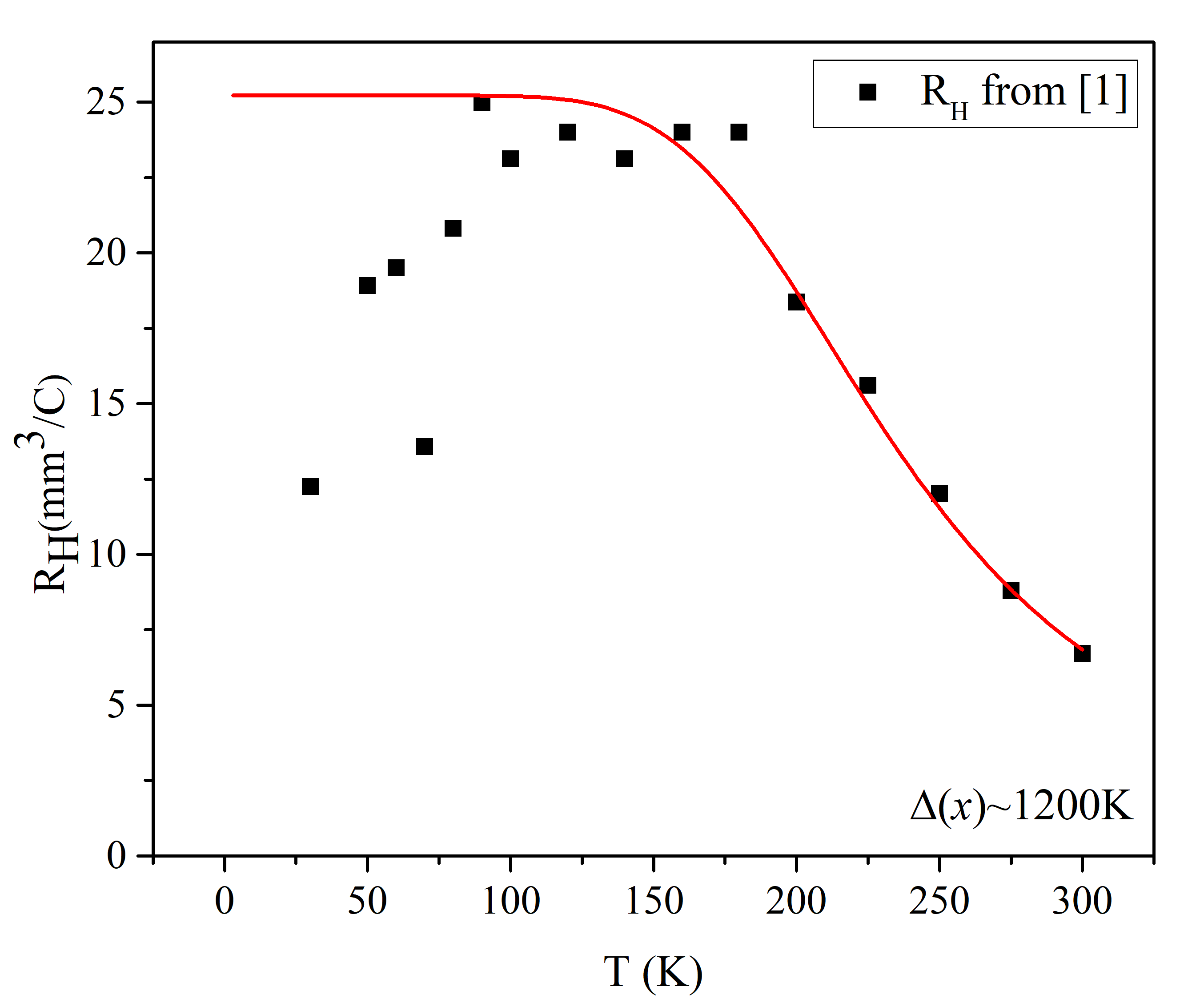}
        \caption{$x$=0.036}
    \end{subfigure}%
    \hfill
    \begin{subfigure}[b]{0.5\columnwidth}
        \includegraphics[width=\textwidth]{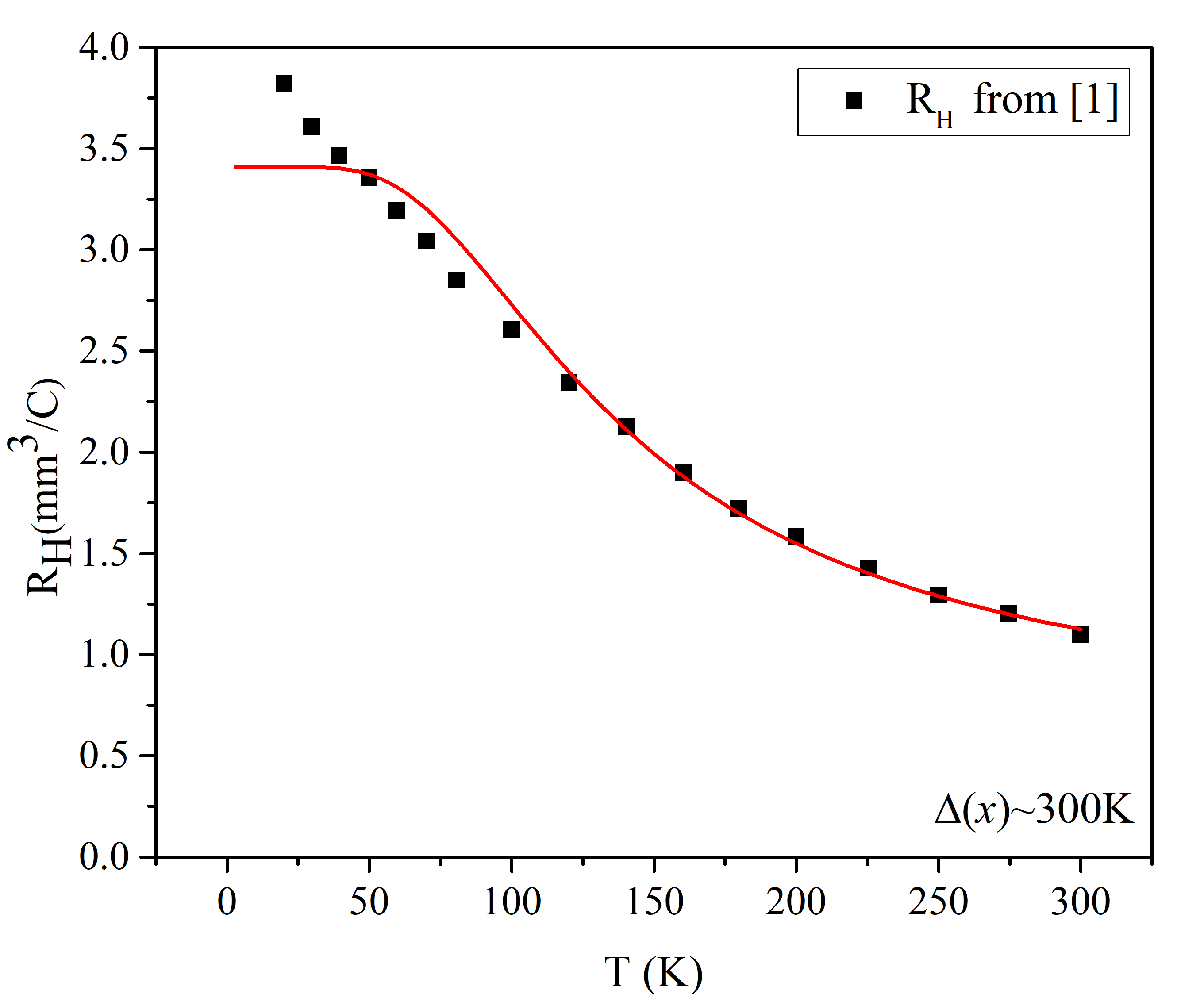}
        \caption{$x$=0.124}
    \end{subfigure}
  \centering
  \caption{\justifying The temperature dependence of Hall coefficients of \(Sr_{2-x}La_{x}IrO_{4}\) for doping concentrations (a) \(x=0.036\) and (b) \(x=0.124\) from the GTTA model (red line) along with experimental \(R_{H}(T)\) from \cite{hsu2023carrier}.}
  \label{fig:2}
\end{figure}

The Hall-effect data for the system was measured very recently by Hsu et al \cite{hsu2023carrier}. The evolution of Hall-coefficient $R_{H}$ with temperature for doping concentrations 0.122, 0.124, 0.138, 0.174, 0.186 and 0.196 is reported. From Fig.1 of \cite{hsu2023carrier}, a qualitative change in the $R_{H}(T)$ is observed for \(x>0.14\) indicating some sort of a crossover. We have extracted the data for of the $R_{H}$ from Fig.1 of \cite{hsu2023carrier} to investigate the nature of PG phase crossover for different La-doping concentrations. Since a crossover is observed near doping \(x\backsim0.16\) only the data for \(x<0.16\) is considered for the present analysis. \(n_{Hall}(x,T)\) is calculated from the extracted $R_{H}(T)$ data \footnote{Before we apply GTTA model to \(Sr_{2-x}La_{x}IrO_{4}\), we reproduce their results \cite{PhysRevLett.97.247003} for cuprates. From the Hall-effect data for $La_{2-x}Sr_{x}CuO_{4}$ \cite{ono2007strong,ando2004evolution}, the carrier density \(n_{hall}\) is calculated using the relation \(R_{H}=\frac{1}{n_{Hall}e}\). The extracted values of \(n_{0}(x), n_{1}\) and $\Delta(x)$ thus obtained are in agreement with values reported by Gor’kov and Teitel’baum. \(n_{0}(x)\), for very low doping (\(x<0.07\)) increases as \(n_{0}(x)\simeq x\) and then non-linearly increases for larger doping \(x\) values as shown in supplementary information. \(n_{1}(x)\) is almost constant \((\simeq2.8)\) and abruptly decreases for large doping (\(x>0.18\)). Doping dependence of $\Delta(x)$ is also shown in the supplementary information. $\Delta(x)$ decreases approximately linearly from $\sim1000K$ at around \(x\simeq0.04\) to $\sim100K$ at \(x\simeq0.20\). Our results agree with the results obtained by Gor’kov and Teitel’baum \cite{PhysRevLett.97.247003}. Thus it benchmarks (or validates) our numerical method.}. Using Eq. 2 the values of \(n_{0}(x), n_{1}\) and $\Delta(x)$ for the mentioned doping concentrations are obtained. The doping dependence $\Delta(x)$ is displayed in Fig. \ref{fig:1}, which is the central result of this work. The $R_{H}(T)$ data for doping concentrations 0.036, 0.08 and 0.10 are obtained via personal communication with the authors of \cite{hsu2023carrier}. The GTTA fitted model along with the reported experimental values of $R_{H}(T)$ for x=0.036 is shown in Fig. \ref{fig:2}(a) for a representation purpose. It is clear that the GTTA model fits exceptionally well for \(T>100 K\) as for low temperatures, weak localization effects comes into picture. For this doping concentration, the value of $\Delta(x)$ reaches $\sim 1200 K$, very similar to that observed in $La_{2-x}Sr_{x}CuO_{4}$. Fig. \ref{fig:2}(b) shows the plot of Hall coefficients vs temperature obtained from the GTTA model for an intermediate doping (\(x=0.124\)). For x=0.124, $\Delta(x)$ is 300 K. The temperature dependence of the Hall-coefficients obtained from GTTA model for all the other doping concentrations are included in the supplementary information. An excellent fitting of GTTA model to the experimental Hall-effect data is seen.

\begin{figure}[h]
  \centering
  \includegraphics[width=1.0\linewidth]{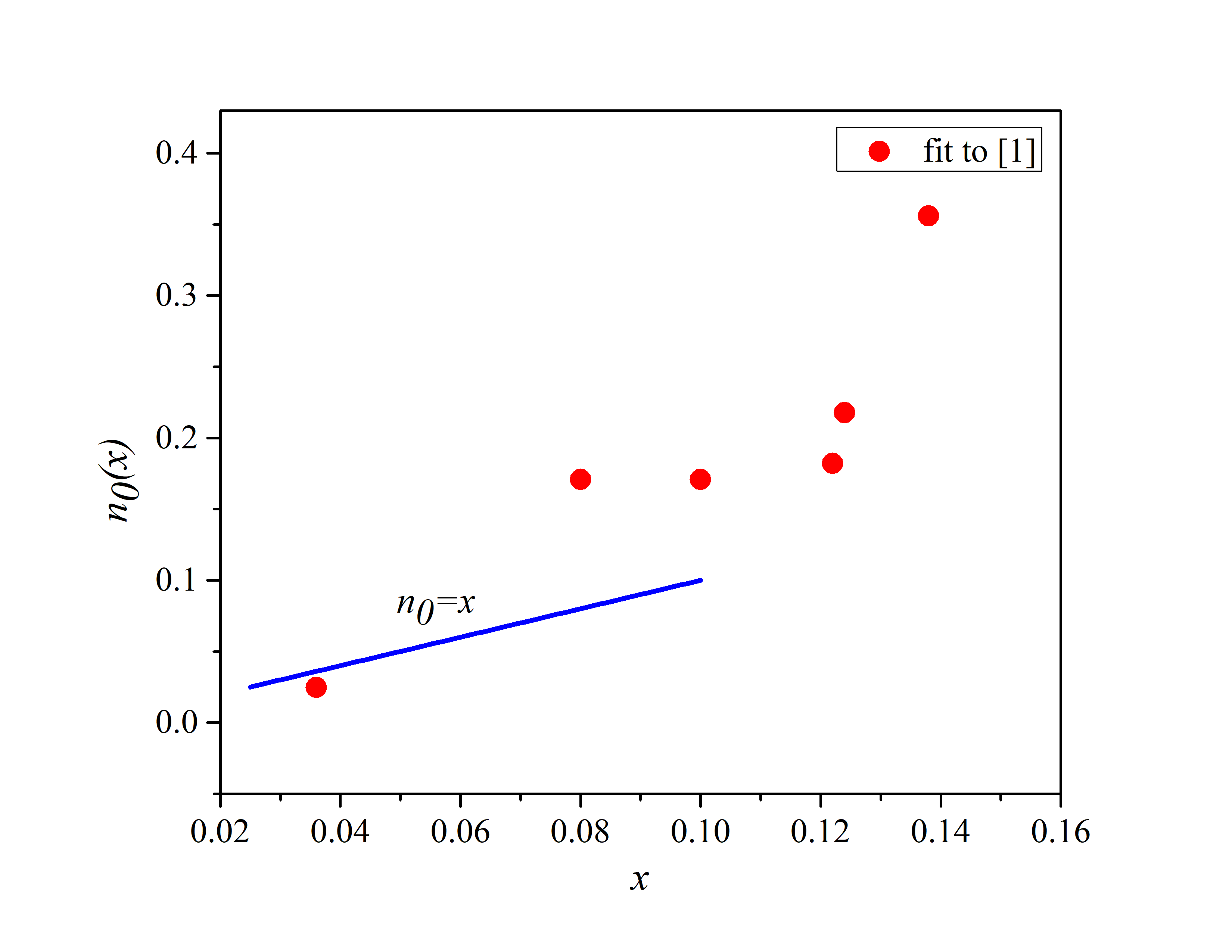}
  \caption{\justifying Plot of \(n_{0}(x)\) as a function of doping calculated from GTTA model using the experimental Hall effect data \cite{hsu2023carrier} for \(Sr_{2-x}La_{x}IrO_{4}\).}
  \label{fig:3}
\end{figure}

Fig. \ref{fig:3} shows the doping dependence of \(n_{0}(x)\). \(n_{0}(x)\) increases with increase in doping concentration. The \(n_{0}(x)\) value for \(x=0.036\) falls near the \(n_{0}(x)\simeq x\) line. For larger \(x\) (\(x>0.036\)), \(n_{0}(x)\) deviates considerably from linear behaviour \(n_{0}(x)=x\) and shoots up for larger \(x\) (\(x\simeq0.14\)) which is the signature of $n_h \sim x$ to $n_h \sim 1+x$ transition. This transition is discussed in detail in \cite{hsu2023carrier}. The doping dependence of (\(n_{0}(x)\)) for \(Sr_{2-x}La_{x}IrO_{4}\) is similar to that observed for $La_{2-x}Sr_{x}CuO_{4}$ \cite{badoux2016change}. \(n_{1}\) is roughly constant (\(n_{1}\sim3.8\)) for low doping (till \(x=0.10\)) after which it abruptly decreases.   

Coming back to Fig. \ref{fig:1}, $\Delta(x)$ starts at a very high temperature (\(\sim1200 K\)) near the Mott insulating phase for low doping (\(x\simeq0.036\)) after which it decreases roughly linearly with increase in electron doping concentration. $\Delta(x)$ decreases to $\sim300K$ at  \(x\simeq0.14\) which is near the predicted quantum critical doping concentration. Data between $x=0.14$ and $x=0.16$ is not available. Regarding the physical meaning of $\Delta(x)$, Gor’kov and Teitel’baum suggested that these thermal activation contributions to the charge carriers come from the activation of electrons from the states near the anti-nodal region \((0,\pm\pi)\) to the nodal \((\pi,\pi)\) region which are separated by an effective energy gap $\Delta(x)$ \cite{PhysRevLett.97.247003}. They validated this interpretation of $\Delta(x)$ by comparing their values with ARPES results showing a strong agreement  \cite{yoshida2003metallic}. Thus, we compare values of $\Delta(x)$ with the PG values of \(Sr_{2-x}La_{x}IrO_{4}\) reported in literature (displayed in Fig.1) obtained by different experimental techniques. In Fig.1 we display two points representing the magnitude of PG for \(x\simeq0.10\) and \(x\simeq0.134\) (red solid circles). The point with \(x\simeq0.10\) corresponds to the PG magnitude along the anti-nodal point obtained through ARPES measurements. The PG magnitude, ($2\Delta_{PG}\simeq50meV$) was extracted from Fig. 3e of \cite{de2015collapse}. The other point corresponds to the PG magnitude $2\Delta_{PG}\sim17meV$ for doping concentration $x\simeq0.134$ measured by Seo et al. using infrared spectroscopy \cite{seo2017infrared}. Thus, the extracted values of $\Delta(x)$ from the Hall effect data using GTTA model are in agreement with the PG signatures already reported in literature for \(Sr_{2-x}La_{x}IrO_{4}\). It would be better if more such experiments are done for the low doping regime so as to more precisely determine the PG boundary.  

The pseudogap data points at low doping such as at $x=0.04$ actually corresponds to a special phase of underdoped \(Sr_{2-x}La_{x}IrO_{4}\). In  reference\cite{Battisti_2016} it is found that underdoped phase of this system consists of an inhomogeneous and phase separated state, more precisely, puddles of pseudogap phase nucleates around dopant atoms forming an inhomogeneous mixture. In other words Mott state and PG state coexist on a mesoscopic length scale. Thus the PG obtained from GTTA analysis in the underdoped regime should be thought of as an average gap. 

\section{Fermi Arcs}
The phenomenon of "Fermi arcs" is observed in cuprates \cite{kondo2013formation}. It is also observed in this iridate system \cite{peng2022electronic}. In cuprates at very low doping PG is of d-wave symmetry which depects a node near $(\pi,\pi)$ direction. At higher dopings and higher temperatures these nodes extend to form an arc lie feature in ARPES spectra around the nodal direction \cite{kondo2013formation}. These features are called "Fermi arcs". 

Using the free electron model for 2-dimensional materials (similar to the approach used by Gor'kov and Teitel'baum \cite{gor2014two}), the relation between number of charge carriers per unit area (N) and the arclength of Fermi arc measured in terms of arc angle $\Delta\phi$ is,
\begin{equation}
N=\frac{1}{2\pi^{2}}\int_{0}^{\Delta\phi}\,d\theta\int_{0}^{k_{F}} k\,dk.
\end{equation}
Thus,
\begin{equation}
    N=\frac{k_{F}^{2}}{4\pi^{2}}\Delta\phi,
    \label{GTTA1}
\end{equation}
where, \(k_{F}\) is the magnitude of Fermi wave vector (in x-y plane). N can be calculated from carrier density (\(n_{Hall}\)) of the material. Thus Eq. 4 can be rewritten as follows:

\begin{equation}
    \Delta\phi=\frac{4\pi^{2}(n_{Hall}c)}{nk_{F}^{2}},
    \label{GTTA2}
\end{equation}

where, c is the lattice constant along the Ir-O-Sr direction i.e. \(25.96\AA\) \cite{crawford1994structural} and $n$ is the number of Ir-O planes in a unit cell of \(Sr_{2}IrO{4}\). Since only one quadrant of Brillouin zone is usually displayed reporting ARPES measurements, we divide $\Delta\phi$ calculated from Eq. \ref{GTTA2} by \ref{GTTA1} to represent the "Fermi arcs".\

\begin{figure}[h]

    \begin{subfigure}[b]{0.5\columnwidth}
        \includegraphics[width=\textwidth]{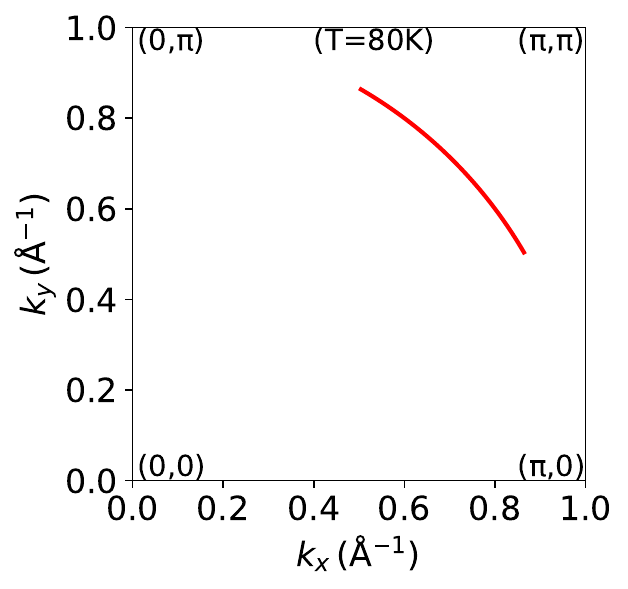}
        \caption{T=80K}
    \end{subfigure}%
    \hfill
    \begin{subfigure}[b]{0.5\columnwidth}
        \includegraphics[width=\textwidth]{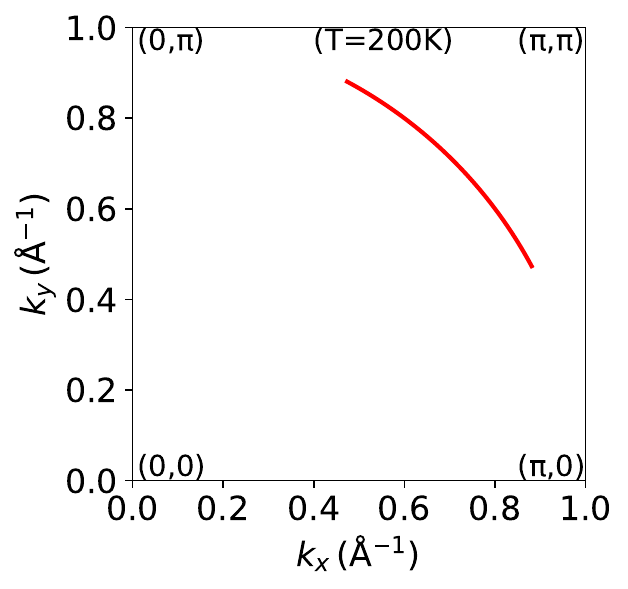}
        \caption{T=200K}
    \end{subfigure}
    \caption{\justifying Evolution of "Fermi arcs" with doping concentration $x=0.08$ at temperatures (a) 80K and (b) 200K}
    \label{fermi arcs}

\end{figure}
The evolution of the "Fermi arcs" $\Delta\phi$ with respect to 
temperature is obtained from Eq. 5 and their schematic plots are displayed in Fig. \ref{fermi arcs}. The magnitude of Fermi wave vector \(k_{F}\) is extracted from Fig-1 of \cite{kim2016observation}. The values of \(n_{Hall}(x,T)\) obtained from the GTTA model are used in Eq. 6 to obtain $\Delta\phi$. The "Fermi arcs", for a given doping concentration expands, i.e. $\Delta\phi$ increases, as the temperature increases. 
Thus with increase in temperature the extension of the gap observed at the anti-nodal region is suppressed for a given doping. Kim et al have reported the "Fermi arcs" obtained by ARPES analysis corresponding to $7\%$ electron-doped  \(Sr_{2}IrO{4}\) in Fig 1 of \cite{kim2016observation} for temperatures 10K and 70K. From Fig 1 of \cite{kim2016observation}, the "Fermi arc" length $\Delta\phi$ for $7\%$ electron-doping at 70K is $\Delta\phi\sim36^{\circ}$ (by taking the angle between points having intensity half to that at nodal point). For doping concentration $x=0.08$ (for which we have the data, and which is close to 7 percent electron doping), the calculated Fermi arc length $\Delta\phi$ at temperature 80K is $\Delta\phi\simeq30^{\circ}$ (Fig. \ref{fermi arcs}). Thus, there is a qualitative agreement between "Fermi arcs" $\Delta\phi$ calculated using Eq. 5, wherein \(n_{Hall}(x,T)\) values are computed using GTTA model, and the reported results \cite{kim2016observation}.    

\section{\label{sec:level1}Conclusions}
\begin{figure}[h] 
  \centering
  \includegraphics[width=1.0 \linewidth]{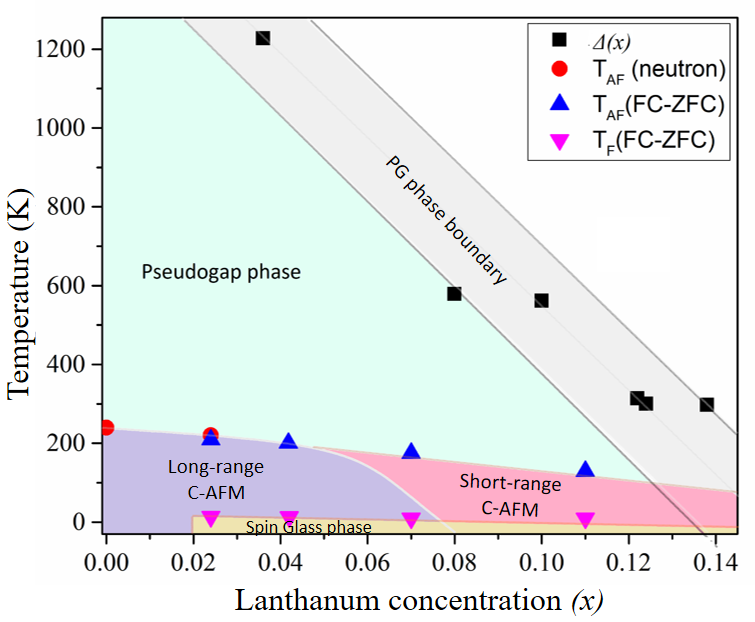}
  \caption{\justifying Updated phase diagram of \(Sr_{2-x}La_{x}IrO_{4}\). The black solid square denotes $\Delta(x)$ obtained from the GTTA model in this work. All the points corresponding to C-AFM transitions is extracted from Fig. 12 of \cite{chen2015influence}. The boundaries between long-range and short-range C-AFM order as suggested in \cite{chen2015influence} is shown with long-range C-AFM order extending upto $x=0.08$ (purple). The proposed region of PG phase boundary for \(Sr_{2-x}La_{x}IrO_{4}\) is marked (in grey). Future experiments and analysis will refine this PG phase boundary.}
  \label{fig:5}
\end{figure}

We have applied the very successful GTTA model to \(Sr_{2-x}La_{x}IrO_{4}\) system and extracted its PG boundary (crossover) using the Hall-effect data \cite{hsu2023carrier}. It is observed that PG opens at very high temperatures ($\sim$ 1200 K) near low doping (\(x\simeq0.036\)) and then it decreases approximately linearly as doping is increased. As displayed in Fig. 1, the PG boundary when extrapolated vanishes at around (\(x\simeq0.16\)) which is the critical doping for the pseudogap opening as discovered in \cite{hsu2023carrier}. Our results also agree with some experimental data for PG previously reported in this system \cite{seo2017infrared,de2015collapse}. We note that this doping dependence of PG boundary is very similar to that found in cuprates. We also  schematically show the evolution of the “Fermi arcs” using the carrier density calculated from the GTTA model at temperatures 80K and 200K for a given doping concentration (\(x\simeq0.08\)) which are in qualitative agreement with the reported results \cite{kim2016observation}. We also draw an updated phase diagram for $Sr_{2-x}La_xIrO_4$ in Fig. \ref{fig:5}. The magnetic phase diagram of the material is redrawn using the extracted data from Fig. 12 of \cite{chen2015influence}. On top of the previously known magnetic phase, we find a region of the PG phase boundary for $Sr_{2-x}La_xIrO_4$ in Fig. \ref{fig:5}. This PG phase boundary should be refined through future experiments and analysis of $Sr_{2-x}La_xIrO_4$ system.  

\section{Acknowledgements} 
The authors are grateful to Nigel E. Hussey and Yu-Te Hsu (and their research group) for giving us the Hall-effect data and also for their valuable inputs through personal correspondence. We also thank Yu-Te Hsu for carefully reading the manuscript and suggesting very important corrections. Devarshi Dave is thankful to Physical Research Laboratory, Ahmedabad, India for providing facilities to carry out this research work.    
\newpage
\bibliographystyle{unsrt}
\bibliography{paper-1}

\begin{thebibliography}{10}

\bibitem{hsu2023carrier}
Yu-Te Hsu, Andreas Rydh, Maarten Berben, Caitlin Duffy, Alberto de~la Torre, Robin~S Perry, and Nigel~E Hussey.
\newblock Carrier density crossover and quasiparticle mass enhancement in a doped 5$ d $ mott insulator.
\newblock {\em arXiv:2312.00515}, 2023.

\bibitem{seo2017infrared}
J.~H. Sue, G.~H. Ahn, S.~J. Song, X.~Chen, S.~D. Wilson, and S.~J. Moon.
\newblock Infrared probe of pseudogap in electron-doped sr2iro4.
\newblock {\em Scientific Reports}, 7(1):10494, 2017.

\bibitem{de2015collapse}
Alberto De~La~Torre, S~McKeown Walker, Flavio~Yair Bruno, Sara Ricc{\'o}, Z.~Wang, I~Gutierrez Lezama, Gernot Scheerer, Ga{\'e}tan Giriat, Didier Jaccard, Christophe Berthod, et~al.
\newblock Collapse of the mott gap and emergence of a nodal liquid in lightly doped ${Sr_2IrO_4}$.
\newblock {\em Physical Review Letters}, 115(17):176402, 2015.

\bibitem{imada1998metal}
Masatoshi Imada, Atsushi Fujimori, and Yoshinori Tokura.
\newblock Metal-insulator transitions.
\newblock {\em Reviews of Modern Physics}, 70(4):1039, 1998.

\bibitem{maekawa2004physics}
Sadamichi Maekawa, Takami Tohyama, Stewart~Edward Barnes, Sumio Ishihara, Wataru Koshibae, and Giniyat Khaliullin.
\newblock {\em Physics of Transition Metal Oxides}, volume 144.
\newblock Springer Science \& Business Media, 2004.

\bibitem{cox2010transition}
Paul~Anthony Cox.
\newblock {\em Transition metal oxides: an introduction to their electronic structure and properties}, volume~27.
\newblock Oxford university press, 2010.

\bibitem{anderson2011personal}
Philip~W. Anderson.
\newblock Personal history of my engagement with cuprate superconductivity, 1986--2010.
\newblock {\em International Journal of Modern Physics B}, 25(01):1--39, 2011.

\bibitem{anderson2004physics}
Philip~W. Anderson, P.~A. Lee, M.~Randeria, T.~M. Rice, N.~Trivedi, and F.~C. Zhang.
\newblock The physics behind high-temperature superconducting cuprates: the ‘plain vanilla’version of rvb.
\newblock {\em Journal of Physics: Condensed Matter}, 16(24):R755, 2004.

\bibitem{lee2006doping}
Patrick~A. Lee, Naoto Nagaosa, and Xiao-Gang Wen.
\newblock Doping a mott insulator: Physics of high-temperature superconductivity.
\newblock {\em Reviews of Modern Physics}, 78(1):17, 2006.

\bibitem{barivsic2013universal}
Neven Bari{\v{s}}i{\'c}, Mun~K. Chan, Yuan Li, Guichuan Yu, Xudong Zhao, Martin Dressel, Ana Smontara, and Martin Greven.
\newblock Universal sheet resistance and revised phase diagram of the cuprate high-temperature superconductors.
\newblock {\em Proceedings of the National Academy of Sciences}, 110(30):12235--12240, 2013.

\bibitem{proust2019remarkable}
Cyril Proust and Louis Taillefer.
\newblock The remarkable underlying ground states of cuprate superconductors.
\newblock {\em Annual Review of Condensed Matter Physics}, 10:409--429, 2019.

\bibitem{singh2021leading}
Navinder Singh.
\newblock Leading theories of the cuprate superconductivity: A critique.
\newblock {\em Physica C: Superconductivity and its Applications}, 580:1353782, 2021.

\bibitem{timusk1999pseudogap}
Tom Timusk and Bryan Statt.
\newblock The pseudogap in high-temperature superconductors: an experimental survey.
\newblock {\em Reports on Progress in Physics}, 62(1):61, 1999.

\bibitem{ayres2022superfluid}
Jake Ayres, Mikhail~I. Katsnelson, and Nigel~E. Hussey.
\newblock Superfluid density and two-component conductivity in hole-doped cuprates.
\newblock {\em Frontiers in Physics}, 10:1021462, 2022.

\bibitem{norman2005pseudogap}
Michael~R. Norman, D.~Pines, and C.~Kallin.
\newblock The pseudogap: friend or foe of high ${T}_c$?
\newblock {\em Advances in Physics}, 54(8):715--733, 2005.

\bibitem{tallon2001doping}
Jeffery~L. Tallon and J.~W. Loram.
\newblock The doping dependence of ${T}^*$- what is the real high-${T}_c$ phase diagram?
\newblock {\em Physica C: Superconductivity}, 349(1-2):53--68, 2001.

\bibitem{tallon2023thermodynamics}
Jeffery~L. Tallon and James~G. Storey.
\newblock Thermodynamics and the pairon model for cuprates.
\newblock {\em Physical Review B}, 107(5):054507, 2023.

\bibitem{tallon2020locating}
Jeffery~L. Tallon, James~G. Storey, John~R. Cooper, and John~W. Loram.
\newblock Locating the pseudogap closing point in cuprate superconductors: Absence of entrant or reentrant behavior.
\newblock {\em Physical Review B}, 101(17):174512, 2020.

\bibitem{taillefer2016change}
Louis Taillefer, Sven Badoux, Gael Grissonnanche, Nicolas Doiron-Leyraud, Wojciech Tabis, Francis Laliberte, Baptiste Vignolle, David Vignolles, Jerome Beard, Cyril Proust, et~al.
\newblock Change of carrier density at the pseudogap critical point of a cuprate superconductor.
\newblock In {\em APS March Meeting Abstracts}, volume 2016, pages K8--013, 2016.

\bibitem{gor2015two}
Lev~P. Gor'kov and Gregory~B. Teitel'Baum.
\newblock Two-component energy spectrum of cuprates in the pseudogap phase and its evolution with temperature and at charge ordering.
\newblock {\em Scientific Reports}, 5(1):8524, 2015.

\bibitem{taillefer2010scattering}
Louis Taillefer.
\newblock Scattering and pairing in cuprate superconductors.
\newblock {\em Annu. Rev. Condens. Matter Phys.}, 1(1):51--70, 2010.

\bibitem{michon2019thermodynamic}
B.~Michon, C.~Girod, Sven Badoux, J.~Ka{\v{c}}mar{\v{c}}{\'\i}k, Qianli Ma, Mirela Dragomir, H.~A. Dabkowska, B.~D. Gaulin, J.~S. Zhou, S.~Pyon, et~al.
\newblock Thermodynamic signatures of quantum criticality in cuprate superconductors.
\newblock {\em Nature}, 567(7747):218--222, 2019.

\bibitem{collignon2017fermi}
C.~Collignon, S.~Badoux, S.~A.~A. Afshar, B.~Michon, F.~Lalibert{\'e}, O.~Cyr-Choini{\`e}re, J.~S. Zhou, S.~Licciardello, S.~Wiedmann, N.~Doiron-Leyraud, et~al.
\newblock Fermi-surface transformation across the pseudogap critical point of the cuprate superconductor ${La}_{1.6-x}{Nd}_0.4{Sr}_x{CuO}_4$.
\newblock {\em Physical Review B}, 95(22):224517, 2017.

\bibitem{bertinshaw2019square}
Joel Bertinshaw, Yeong~Kwan Kim, Giniyat Khaliullin, and B.~J. Kim.
\newblock Square lattice iridates.
\newblock {\em Annual Review of Condensed Matter Physics}, 10:315--336, 2019.

\bibitem{chen2015influence}
Xiang Chen, Tom Hogan, D.~Walkup, Wenwen Zhou, M.~Pokharel, Mengliang Yao, Wei Tian, Thomas~Z. Ward, Yang Zhao, D.~Parshall, et~al.
\newblock Influence of electron doping on the ground state of $({Sr}_{1-x}{La}_x)_2{Ir}{O}_4$.
\newblock {\em Physical Review B}, 92(7):075125, 2015.

\bibitem{kim2014fermi}
Yeong~Kwan Kim, O.~Krupin, J.~D. Denlinger, A.~Bostwick, E.~Rotenberg, Q.~Zhao, J.~F. Mitchell, J.~W. Allen, and B.~J. Kim.
\newblock Fermi arcs in a doped pseudospin-1/2 heisenberg antiferromagnet.
\newblock {\em Science}, 345(6193):187--190, 2014.

\bibitem{kim2016observation}
Yeong~Kwan Kim, N.~H. Sung, J.~D. Denlinger, and B.~J. Kim.
\newblock Observation of ad-wave gap in electron-doped ${S}r_2{IrO}_4$.
\newblock {\em Nature Physics}, 12(1):37--41, 2016.

\bibitem{li2015tuning}
Ming-Ying Li, Zheng-Tai Liu, Hai-Feng Yang, Jia-Lin Zhao, Qi~Yao, Cong-Cong Fan, Ji-Shan Liu, Bo~Gao, Da-Wei Shen, and Xiao-Ming Xie.
\newblock Tuning the electronic structure of ${Sr_2IrO_4}$ thin films by bulk electronic doping using molecular beam epitaxy.
\newblock {\em Chinese Physics Letters}, 32(5):057402, 2015.

\bibitem{PhysRevLett.97.247003}
Lev~P. Gor'kov and Gregory~B. Teitel'baum.
\newblock Interplay of externally doped and thermally activated holes in ${\mathrm{la}}_{2\ensuremath{-}x}{\mathrm{sr}}_{x}{\mathrm{cuo}}_{4}$ and their impact on the pseudogap crossover.
\newblock {\em Physical Review Letters}, 97:247003, Dec 2006.

\bibitem{ono2007strong}
S.~Ono, Seiki Komiya, and Yoichi Ando.
\newblock Strong charge fluctuations manifested in the high-temperature hall coefficient of high-${T}_c$ cuprates.
\newblock {\em Physical Review B}, 75(2):024515, 2007.

\bibitem{ando2004evolution}
Yoichi Ando, Y~Kurita, Seiki Komiya, S.~Ono, and Kouji Segawa.
\newblock Evolution of the hall coefficient and the peculiar electronic structure of the cuprate superconductors.
\newblock {\em Physical Review Letters}, 92(19):197001, 2004.

\bibitem{yoshida2003metallic}
T.~Yoshida, X.~J. Zhou, T.~Sasagawa, W.~L. Yang, P.~V. Bogdanov, A.~Lanzara, Z.~Hussain, T.~Mizokawa, A.~Fujimori, H.~Eisaki, et~al.
\newblock Metallic behavior of lightly doped ${La}_{2-x}{Sr}_x{CuO}_4$ with a fermi surface forming an arc.
\newblock {\em Physical Review Letters}, 91(2):027001, 2003.

\bibitem{yoshida2006systematic}
T.~Yoshida, X.~J. Zhou, K.~Tanaka, W.~L. Yang, Z.~Hussain, Z.~X. Shen, A.~Fujimori, S.~Sahrakorpi, M.~Lindroos, R.~S. Markiewicz, et~al.
\newblock Systematic doping evolution of the underlying fermi surface of \({La}_{2-x}{Sr}_x{CuO}_4\).
\newblock {\em Physical Review B}, 74(22):224510, 2006.

\bibitem{Note1}
Before we apply GTTA model to \(Sr_{2-x}La_{x}IrO_{4}\), we reproduce their results \cite {PhysRevLett.97.247003} for cuprates. From the Hall-effect data for $La_{2-x}Sr_{x}CuO_{4}$ \cite {ono2007strong,ando2004evolution}, the carrier density \(n_{hall}\) is calculated using the relation \(R_{H}=\protect \frac {1}{n_{Hall}e}\). The extracted values of \(n_{0}(x), n_{1}\) and $\Delta (x)$ thus obtained are in agreement with values reported by Gor’kov and Teitel’baum. \(n_{0}(x)\), for very low doping (\(x<0.07\)) increases as \(n_{0}(x)\simeq x\) and then non-linearly increases for larger doping \(x\) values as shown in supplementary information. \(n_{1}(x)\) is almost constant \((\simeq 2.8)\) and abruptly decreases for large doping (\(x>0.18\)). Doping dependence of $\Delta (x)$ is also shown in the supplementary information. $\Delta (x)$ decreases approximately linearly from $\sim 1000K$ at around \(x\simeq 0.04\) to $\sim 100K$ at \(x\simeq 0.20\). Our results agree with the results obtained by
  Gor’kov and Teitel’baum \cite {PhysRevLett.97.247003}. Thus it benchmarks (or validates) our numerical method.

\bibitem{badoux2016change}
Sven Badoux, Wojciech Tabis, F.~Lalibert{\'e}, G.~Grissonnanche, Baptiste Vignolle, David Vignolles, Jerome B{\'e}ard, D.~A. Bonn, W.~N. Hardy, R.~Liang, et~al.
\newblock Change of carrier density at the pseudogap critical point of a cuprate superconductor.
\newblock {\em Nature}, 531(7593):210--214, 2016.

\bibitem{Battisti_2016}
I.~Battisti, K.~M. Bastiaans, V.~Fedoseev, A.~de~la Torre, N.~Iliopoulos, A.~Tamai, E.~C. Hunter, R. S. Perry, J.~Zaanen, F.~Baumberger, and M.~P. Allan.
\newblock Universality of pseudogap and emergent order in lightly doped mott insulators.
\newblock {\em Nature Physics}, 13(1):21–25, September 2016.

\bibitem{kondo2013formation}
Takeshi Kondo, Ari~D. Palczewski, Yoichiro Hamaya, Tsunehiro Takeuchi, J.~S. Wen, Z.~J. Xu, Genda Gu, and Adam Kaminski.
\newblock Formation of gapless fermi arcs and fingerprints of order in the pseudogap state of cuprate superconductors.
\newblock {\em Physical Review Letters}, 111(15):157003, 2013.

\bibitem{peng2022electronic}
Shuting Peng, Christopher Lane, Yong Hu, Mingyao Guo, Xiang Chen, Zeliang Sun, Makoto Hashimoto, Donghui Lu, Zhi-Xun Shen, Tao Wu, et~al.
\newblock Electronic nature of the pseudogap in electron-doped ${Sr_2IrO_4}$.
\newblock {\em npj Quantum Materials}, 7(1):58, 2022.

\bibitem{gor2014two}
Lev~P. Gor’kov and Gregory~B. Teitel’baum.
\newblock Two regimes in conductivity and the hall coefficient of underdoped cuprates in strong magnetic fields.
\newblock {\em Journal of Physics: Condensed Matter}, 26(4):042202, 2014.

\bibitem{crawford1994structural}
M.~K. Crawford, M.~A. Subramanian, R.~L. Harlow, J.~A. Fernandez-Baca, Z.~R. Wang, and D.~C. Johnston.
\newblock Structural and magnetic studies of ${Sr_2IrO_4}$.
\newblock {\em Physical Review B}, 49(13):9198, 1994.

\end{thebibliography}
\end{document}